\documentclass[12pt,preprint]{aastex}
\def\jcap{JCAP}
\def\beq{\begin{equation}}
\def\eeq{\end{equation}}
\def\ben{\begin{eqnarray}}
\def\een{\end{eqnarray}}

\def\lcdm{\Lambda{\rm CDM}}
\def\nucdm{\nu{\rm CDM}}

\def\frs{{\rm fR6}}
\def\frss{{\rm fR6+0.06}\,{\rm eV}}
\def\frff{{\rm fR5+0.15}\,{\rm eV}}

\def\munit{\,h^{-1}\! M_{\odot}}

\def\ev{\,{\rm eV}}

\def\scal{\vert f_{R0}\vert}
\def\mnu{\sum m_{\nu}}
\def\vr{v_{r}(r)}
\def\tr{\tilde{r}}
\def\tv{\tilde{v}_{r}}
\usepackage{color}

\begin{document}
\title{Combined Effects of $f(R)$ Gravity and Massive Neutrinos on the Turn-Around Radii of Dark Matter Halos}
\author{Jounghun Lee\altaffilmark{1}, Marco Baldi\altaffilmark{2,3,4}}
\altaffiltext{1}{Astronomy Program, Department of Physics and Astronomy, FPRD, 
Seoul National University, Seoul 08826, Korea \email{jounghun@astro.snu.ac.kr}}
\altaffiltext{2}{Dipartimento di Fisica e Astronomia, Alma Mater Studiorum Universit\`a di Bologna, viale Berti Pichat, 
6/2, I-40127 Bologna, Italy}
\altaffiltext{3}{INAF - Osservatorio Astronomico di Bologna, via Ranzani 1, I-40127 Bologna, Italy}
\altaffiltext{4}{INFN - Sezione di Bologna, viale Berti Pichat 6/2, I-40127 Bologna, Italy}
\begin{abstract}
We present a new statistics based on the turn-around radii of cluster halos to break the dark sector degeneracy between the 
$\lcdm$ model and the alternative ones with $f(R)$ gravity and massive neutrinos ($\nu$) characterized by 
the strength of the fifth force, $\scal$, and the total neutrino mass, $M_{\nu}$. Analyzing the rockstar halo catalogs at the present epoch from 
the {\small DUSTGRAIN}-{pathfinder} $N$-body simulations performed for four different cosmologies, namely, 
$\lcdm$ ($\scal=0$, $\mnu=0.0\ev$), fR6 ($\scal=10^{-6}$, $\mnu=0.0\ev$), fR6+$0.06\ev$ ($\scal=10^{-6}$, $\mnu=0.06\ev$) and 
fR5+$0.15\ev$ ($\scal=10^{-5}$, $\mnu=0.15\ev$), which are known to yield very similar conventional statistics to one another.  
For each model, we select those cluster halos which do not neighbor any other larger halos in their bound zones and construct their bound-zone 
peculiar velocity profiles at $z=0$. Then, we determine the radial distance of each  selected halo at which the bound-zone velocity becomes equal to 
the recession speed of the Hubble flow as its turn around radius, and evaluate the cumulative probability distribution of the ratios of the turn-around 
radii to the virial counterparts, $P(r_{t}/r_{v}\ge \alpha)$.  
The degeneracy between the fR6 and fR5+$0.15\ev$ models is found to be readily broken by the $10\sigma_{\Delta P}$ difference in the value of $P(\alpha=4)$, 
while the $3.2\sigma_{\Delta P}$ difference between the $\lcdm$ and fR6+$0.06\ev$ models is detected in the value of $P(\alpha=8.5)$. 
It is also found that the four models yield smaller differences in $P(\alpha)$ at higher redshifts. 
\end{abstract}
\section{Introduction}\label{sec:intro}

The turn-around radius of a dark matter (DM) halo is a characteristic distance scale at which the velocity field around the halo has a vanishingly small value in 
the radial direction due to the complete counter-balance between its inward gravity  and the outward repulsion of the Hubble flow. 
Even though the turn-around radius is a property of a highly nonlinear structure, its value can in principle be theoretically 
predictable from the first principles as far as the halo forms through the spherically symmetric gravitational collapse process \citep{PT14,pav-etal14}. 
This advantageous aspect of the turn-around radius has motivated many authors to examine its potential as a probe of cosmology. 
For example, \citet{PT14} analytically evaluated the upper limit on the turn-around radii for the standard $\lcdm$ cosmology, where the gravitational law is described 
by Einstein's general relativity (GR), the present acceleration of the universe is driven by the cosmological constant ($\Lambda$) with equation of state $w=-1$, 
and the most dominant matter content is the collisionless cold DM (CDM) particles having negligibly low speed at the moment of their decoupling. 

What \citet{PT14} proved was that the spherical upper limit on the turn-around radii sensitively depends on the amount of $\Lambda$  \citep[see also][]{pav-etal14,BT17} 
and thus that a bound violation, if observed to occur, could in principle challenge the $\lcdm$ cosmology. 
Here, a bound violation is a term coined by \citet{PT14} to describe an event of observing a cosmic structure whose turn-around radius exceeding the analytically found {\it spherical} 
upper limit of the $\lcdm$ cosmology. 
Later, \citet{lop-etal18} theoretically proved that the upper limit of the turn-around radii can be used to detect the presence of modified gravity \citep[MG,][for a review]{mgreview} which 
has an effect of significantly increasing the turn-around radii \citep[see also][]{lop-etal19}. 

The aforementioned theoretical works were based on the simple top-hat spherical dynamics, from which the real gravitational dynamics is in fact well known to depart 
\citep[e.g.,][]{BM96}. To take into account the non-spherical nature of gravitational collapse for the determination of the turn-around radii and their upper limit, 
the numerical experiments had to be employed \citep{PT14}. For instance, \citet{LY16} used a high-resolution N-body simulation to measure the turn-around 
radii of DM halos located in the cosmic web \citep{web96} and demonstrated that the anisotropic merging along the filamentary structures has an effect of enlarging 
the turn-around radii. Their result implied that the occurrence of a bound violation is not unconditionally prohibited but occasionally possible even in the $\lcdm$ cosmology since the 
numerically determined non-spherical upper limit on the turn-around radii turned out to be higher than the analytically predicted spherical limit  \citep[see also][]{BT21,far-etal21,GF21}.

Nevertheless, the usefulness of the turn-around radii as a cosmological probe is not necessarily undermined by the fact that their upper limit cannot be treated in a purely 
analytical way. \citet{LL17} numerically found that  it becomes significantly more probable for a bound violation to occur in the presence of MG and thus that the frequency 
of the occurrence of the bound violations should be a powerful test of GR. Their claim was supported by several follow-up works which theoretically proved that the alternative 
cosmologies including quintessence dark energy (DE), scalar-tensor theory, and phantom brane world induce more frequent occurrence of the bound violations 
\citep{BT17,BK17,noj-etal18,lop-etal18,lop-etal19}.

In light of the aforementioned works which disclosed the sensitivity of the turn-around radii especially to the nature of gravity, 
we attempt here to numerically explore if the turn-around radii is capable of discriminating the alternative MG models that have been known to be 
degenerate with the $\lcdm$ cosmology by the conventional statistics such as the linear and nonlinear density power spectra, cluster mass function, halo bias factor 
and redshift distortion effect \citep{bal-etal14,hag-etal19a}.    
For this exploration, our analysis will focus on a particular class of MG models, namely, the $\nucdm$+$f(R)$ gravity model, where the massive neutrinos ($\nu$) with 
non-zero total mass $\mnu$ is present along with CDM and the apparent acceleration of space time at the present epoch is caused by the failure of GR on the 
cosmological scales \citep[see][for a review]{DT10}. 

The gravitational dynamics of this alternative model is dictated by the modified Einstein-Hilbert action in which some specified function, $f(R)$, 
substitutes for the Ricci scalar $R$. An additional {\it fifth} force is generated by its extra degree of freedom, $f_{R}\equiv df/dR$, dubbed the {\it scalaron}, 
whose absolute value at the present epoch, $\scal$, quantifies how strong fifth force the $f(R)$ gravity can exert \cite[see][and references therein]{HS07}.
Although the Chameleon shielding mechanism turns off the fifth force in the high-density regions, the overall effect of $f(R)$ gravity alone is to enhance the density growth via 
its fifth force compared with the $\lcdm$ case \citep[see][and references therein]{KW04}.  However, in the presence of massive neutrinos which has an effect of suppressing the 
density growth \citep[e.g., see][]{LP14}, this effect of $f(R)$ gravity can be severely attenuated. It was indeed numerically shown that a proper combination of $\mnu$ with 
$\scal$ can make a $\nucdm$+$f(R)$ gravity model to yield very similar conventional statistics to the $\lcdm$ case \citep{bal-etal14,hag-etal19a}. 

In this Paper, we will provide a numerical evidence supporting that the turn-around radii may break this degeneracy between the $\lcdm$ and $\nucdm$+$f(R)$ 
gravity models. 
The organization of this paper is as follows. In Section \ref{sec:review},  we briefly review the previously developed algorithm for the estimation of the turn-around radii 
of DM halos. In Section \ref{sec:analysis},  we describe the numerical data used for our analysis and explain how well the cumulative probability of the  turn-around radii 
of DM halos differentiate between the $\lcdm$ and $\nucdm$+$f(R)$ gravity models. In Section \ref{sec:con}, we summarize the results and discuss the caveats 
and limitations of our statistics as a cosmological discriminator. 

\section{A Review of the TRE algorithm}\label{sec:review}

The neighborhood around a DM halo is often divided into three distinct sectors called the {\it infall, bound} and {\it Hubble} zones, 
depending on which effect is more dominant between the gravity and the cosmic expansion. The infall (Hubble) zone corresponds to 
the radial distance range, $r\le 2r_{v}$ ($r\ge 10r_{v}$), in which the effect of the gravitational attraction of the halo on the radial components of 
the peculiar velociteis, $v_{r}$, completely surpasses (surrenders) that of the receding Hubble flow, where $r_{v}$ is the halo virial radius. 
Meanwhile, the in-between bound zone corresponds to the region where the two competing forces are so well balanced that $v_{r}$ can be tractable 
in the linear perturbation theory. 

\citet{fal-etal14} showed that the profile of the radial components of the peculiar velocity field in the bound zone, $\vr$, around the cluster halos with virial mass 
$M_{v}\gtrapprox 0.5\times 10^{14}\munit$ has a universal shape, well approximated by the following formula, 
\beq
\label{eqn:vr}
\frac{\vr}{V_c} =  - A\left(\frac{r_{v}}{r}\right)^{n}\, , 
\eeq
where two adjustable parameters, $A$ and $n$, quantify the amplitude and slope of the profile, respectively, and  $V_{c}$ is the circular velocity equivalent 
to $(GM_{v}/r_{v})^{1/2}$. The negative sign in the right-hand side of Equation (\ref{eqn:vr}) indicates that the bound-zone neighbors still feel the net gravitational 
force of the halo. From here on, we let $\vr$ exclusively denote the profile of the radial components of the peculiar velocity field in the bound zone around the 
cluster halos and call it the {\it bound-zone velocity profile}.

\citet{fal-etal14} claimed the {\it universality} of Equation (\ref{eqn:vr}) based on their numerical finding that the stacked bound-zone velocity profiles over 
the cluster halos has a constant slope and amplitude, being almost independent of the cluster masses and redshifts.   
It was also found by \citet{fal-etal14} that not only the stacked ones but also the bound-zone velocity profiles, $\vr$, around individual cluster halos follow well 
the above formula, although the best-fit values of $A$ and $n$ exhibited substantial scatters around the mean values. 
The pioneering work of \citet{fal-etal14} motivated further numerical investigations of the bound-zone velocity profiles around the cluster halos, which all 
confirmed the validity of Equation (\ref{eqn:vr}) for the description of $\vr$ \citep{lee-etal15,lee16,LY16,lee18,han-etal20}.  

\citet{LY16} demonstrated with the help of a N-body simulations that the best-fit values of $A$ and $n$ in Equation (\ref{eqn:vr}) depend on the halo environments and that the 
best-agreements between the numerically obtained $\vr$ and Equation (\ref{eqn:vr}) can be achieved for the case that the cluster halos are located in the relatively low-density 
environments, having no larger neighbor halos in their bound zones. They also showed that even when $\vr$ is constructed not directly from the DM particles but only from the distinct 
neighbor halos in the bound zones, it is well described by Equation (\ref{eqn:vr}), proving the feasibility of the observational application of Equation (\ref{eqn:vr}) to real data.
\citet{lee16} confirmed the universality of Equation (\ref{eqn:vr}), showing that the best-fit values of $A$ and $n$ in Equation (\ref{eqn:vr}) are quite insensitive to the variation of the 
key cosmological parameters, $\sigma_{8}$ and $\Omega_{m}$. It was also found by \citet{lee16} that Equation (\ref{eqn:vr}) is valid even on the lower mass scales corresponding 
to the group-size halos, $5\times 10^{12}\le M_{v}/(\munit)\le 10^{13}$, as far as the halos are located in the isolated regions. 

\citet{lee-etal15} proposed an algorithm based on Equation (\ref{eqn:vr}) to estimate the turn-around radii, $r_{t}$, of the cluster halos, calling it the turn-around radius estimator (TRE). 
By definition, the magnitude of the bound-zone velocity, $v_{r}$, at $r_{t}$ becomes equal to the speed of the Hubble flow, $H_{0}r_{t}$. 
By equation (\ref{eqn:vr}), however, $\vert v_{r}(r_{t})\vert$ is nothing but $A\,\left[r_{v}/r_{t}\right]^{n}$.  Given that $A$ and $n$ have constant values, the turn-around radius of 
a cluster halo can be estimated simply by solving the following equation. 
\beq
\label{eqn:rt}
A\,\left[\frac{r_{v}}{r_{t}}\right]^{n}\ = \frac{H_{\rm 0}r_{t}}{V_{c}}\, .
\eeq
\citet{LL17} applied the TRE to the numerical data from a high resolution N-body simulation and found that the TRE worked better when they used the best-fit values of $A$ and $n$ 
obtained separately for each cluster halo rather than using their constant mean values. 

For the application of the TRE to the real clusters from observations, the critical issue to address was whether or not the values of $A$ and $n$ for the individual clusters could 
be obtained without measuring the bound-zone peculiar velocities.  According to \citet{fal-etal14}, for those cluster halos embedded in cosmic filaments, it is possible to 
construct $\vr$ from limited information only on the redshift space positions of the bound-zone galaxies. Once $\vr$ is constructed for the filament clusters whose virial mass and 
radius are known, then the values of $A$ and $n$ can be readily obtained through fitting of $\vr$ to Equation (\ref{eqn:vr}). \citet{lee18} applied this TRE to the local 
galaxy clusters located in the straight filamentary structures and successfully estimated their turn-around radii, validating its practical usefulness. 
In Section \ref{sec:analysis}, we will apply the TRE to the numerical data for the investigation of the combined effects of $f(R)$ gravity and massive neutrinos on the turn-around radii 
of the cluster halos.

\section{Physical Analysis}\label{sec:analysis}

The {\small DUSTGRAIN}-{\it pathfinder} simulation project aimed at keeping track of $768^{3}$ CDM particles of individual mass $8.1\times 10^{10}\munit$
under the influence of $f(R)$ gravity in the presence of massive neutrinos \citep{gio-etal19} on the periodic box of volume $750^{3}\,h^{-3}$Mpc$^{3}$. 
The {\small MG-GADGET} encoded by \citet{mggadget} was implemented for the computation of the Hu-Sawicki $f(R)$ gravity \citep{HS07}, while the 
incorporation of massive neutrinos was achieved via the particle-based routine programmed by \citet{vie-etal10}. 
See \citet{gio-etal19} and \citet{mggadget} for detailed information on the {\small DUSTGRAIN}-{\it pathfinder} project and {\small MG-GADGET} code, 
respectively. Among many $\nu$CDM+$f(R)$ gravity models simulated by the {\small DUSTGRAIN}-{\it pathfinder},  we consider the following three, 
fR6 ($\scal=10^{-6}$, $\mnu=0.0\ev$), $\frss$ ($\scal=10^{-6}$, $\mnu=0.06\ev$),  and $\frff$ with $\scal=10^{-5}$ and $\mnu=0.15\ev$) models. 
As mentioned in Section \ref{sec:intro}, it was shown by the previous works of \citet{bal-etal14} and \citet{hag-etal19a} that the conventional statistics can hardly 
discriminate these three models from the $\Lambda$CDM cosmology, which was also simulated by the {\small DUSTGRAIN}-{\it pathfinder} project, 
setting the initial conditions at the Planck values \citep{planck16}. 

For each of the four models, i.e., the $\lcdm$ and the three $\nucdm$+$f(R)$ gravity cosmologies, 
\citet{lee-etal22} identified the DM halos by applying the Rockstar algorithm \citep{rockstar} to the snapshot data of the {\small DUSTGRAIN}-{\it pathfinder} 
simulations. From the Rockstar halo catalogs at $z=0$,  we extract the cluster-size distinct halos with $M_{v}\ge 4.05\times 10^{13}\munit$ enclosing $500$ 
or more DM particles within their virial radii $r_{v}$.  Among them, we select only those cluster halos which do not neighbor any higher-mass halos in their 
bound zones.  From here on, the selected cluster halos will be referred to as the host halos, for which we find the neighbor halos located in their bound-zones and  
containing $20$ or more DM particles. Then, we compute the radial components of the relative peculiar velocities of the bound-zone neighbors around each host as
\beq
\label{eqn:rel}
v_{r} = \hat{\bf r}\cdot\left({\bf V}_{h}-{\bf V}_{b}\right)\, ,
\eeq
where ${\bf V}_{h}$ and ${\bf V}_{b}$ denote the comoving peculiar velocities of a host halo and its bound-zone neighbor halo, respectively, and $\hat{\bf r}$ 
is the unit vector in the direction from the host halo center to the bound-zone neighbor separated by a distance $r$.  The rescaled bound-zone velocities 
and separation distances, $\tv$ and $\tr$, are defined as $\tilde{v}_{r}\equiv v_{r}/V_{c}$ and $\tr\equiv r/r_{v}$, respectively. From here on, we will call 
$\tv$ the bound-zone velocity profile, dropping the term, "rescaled", unless otherwise stated. 

Breaking up the bound-zone distance range of $2<\tr<10$ into several intervals of equal length, $\Delta\tr=1$, we take the average of $\tv$ over 
those neighbors with $\tr$ falling in each interval to determine the bond-zone velocity profile, $\tv(r)$, for each host. 
Then, we take its ensemble average over all of the hosts to determine the stacked bound-zone velocity profile, $\langle\tv\rangle$, 
the result of which is shown in Figure \ref{fig:vr}. 
As can be seen, the average bound-zone velocity profiles conspicuously differ between the $\frff$ and other three models. 
The former case yields significantly higher values of $\vert\langle\tv\rangle\vert$ in the whole range of $\tr$ than 
the latter case, which must be caused by the difference in the strength of the fifth force between the two cases. 
This result implies that the free streaming of more massive neutrinos present in the $\frff$ model do not severely attenuate 
the effect of it stronger fifth force on the bound-zone velocity profiles.

We fit $\langle\tv\rangle$ to Equation (\ref{eqn:vr}) by simultaneously adjusting $A$ and $n$ with the help of the $\chi^{2}$-minimization 
method for each model. Table \ref{tab:data} lists their best-fit values in the third and fourth columns for the four cosmologies. 
Figure \ref{fig:the} compares the numerically obtained $\langle\tv\rangle$ (black filled circles) to the analytic formula with the best-fit 
values of $A$ and $n$ (red solid lines), revealing the excellent agreements between the numerical and analytical results. This result 
confirms the validity and usefulness of Equation (\ref{eqn:vr}) not only for the $\lcdm$ model but also for the $\nu$CDM+$f(R)$ gravity models.
Figure \ref{fig:chi2} plots the $68\%$, $95\%$ and $99\%$ contours of $\chi^{2}(A,n)$, quantifying how significant the differences in the best-values of 
$A$ and $n$ are between the $\frff$ and the other three models.  

Recall that the $\frff$ model is very similar to the other three ones in the conventional statistics \citep{bal-etal14}. Especially the $\frs$ model has been shown to be almost 
indistinguishable from the $\frff$, as the effect of the stronger fifth force, $\vert f_{R0}\vert = 10^{-5}$ is so severely attenuated by the free streaming massive neutrinos 
with $\mnu=0.15\ev$ that the palpable effect amounts only to that of $\vert f_{R0}\vert = 10^{-6}$.
However, the average bound-zone velocity profile, unlike the aforementioned conventional statistics, is capable of disentangling the effect of the fifth force 
from that of the massive neutrinos, sensitively varying with the former, but not with the latter. 
Yet, given that the other three models still remain  mutually indistinguishable even by $\langle \vr\rangle$, we now investigate if any other statistics based on $\vr$ 
beyond its ensemble average can break the degeneracy among the other three models. Basically, we consider the turn-around radii of the hosts as such a statistics, 
 and estimate them by applying the TRE reviewed in Section \ref{sec:review} to the individual bound-zone velocity profiles.

For each host, we fit the individual bound-zone velocity profile to Equation (\ref{eqn:vr}) and separately determine the best-fit values of $A$ and $n$. While performing 
this fitting procedure, we exclude a small fraction of the hosts whose bound-zone velocity profiles fail to be fitted by Equation (\ref{eqn:vr}) due to the low number 
of their bound-zone neighbors. Table \ref{tab:data} lists the number of the hosts, $N_{h}$, whose $\vr$ matches Equation (\ref{eqn:vr}) for the four cosmologies in the 
second column. Plugging the best-fit values of $A$ and $n$ into Equation (\ref{eqn:rt}) and solving it, we estimate the turn-around radius, $r_{t}$, of each of the 
included hosts. Counting the host halos whose turn-around radii exceed $\alpha r_{v}$ as a function of a dimensionless variable $\alpha$, we obtain the cumulative probability 
$P(r_{t}\ge \alpha r_{v})$. To assess the errors, $\sigma_{P}$, in $P(r_{t}\ge \alpha r_{v})$, we create $10,000$ bootstrap resamples composed of equal number of the hosts 
and obtain the cumulative probabilities from each resample. The one standard scatter among the resamples from the average is taken as $\sigma_{P}$. 

Figure \ref{fig:cpro} plots the cumulative probabilities,  $P(r_{t}\ge \alpha r_{v})$, with the bootstrap errors, $\sigma_{P}$, for the four models. 
As expected, the $\frff$ model yields the most conspicuously different $P(r_{t}\ge \alpha r_{v})$ from the other three models. 
The host halos in the $\frff$ model seem to have much larger turn-around radii than in the other three models, producing $22\%$ higher value of $P(\alpha=4)$, 
corresponding to a $10\sigma_{\Delta P}$ signal, where the uncertainties, $\sigma_{\Delta p}$, associated with the measurement of the difference between the 
cumulative probabilities, are computed through the propagation of the bootstrap errors. 

Regarding the other three models, they yield almost identical values of $P(r_{t}\ge \alpha r_{v})$ up to $\alpha=8$. However, the two models, $\lcdm$ and $\frss$, 
which are in fact indistinguishable by the conventional statistics \citep{bal-etal14,hag-etal19a}, show different behaviors in the limit of $\alpha>8$.  
The difference in the value of $P(\alpha=8.5)$ between the two models is found to be as high as $3.2\sigma_{\Delta P}$ in spite of the strongest degeneracy 
between them, yielding the same value of $\sigma_{8}$ \citep{bal-etal14,hag-etal19a}.  Table \ref{tab:data} lists $P(\alpha=4)$ and $P(\alpha=8.5)$ for the four cosmologies in the fifth and 
sixth columns.  Noting that the $\frss$ model yields a smaller value of $P(\alpha=8.5)$ than the other models and that the $\frs$ model does not show any significant difference in the 
whole range of $\alpha$ from the $\lcdm$ case, we suspect the following: Even though the bound-zone velocity profiles are less sensitive to the presence of massive neutrinos, 
the effect of $f(R)$ gravity combined with massive neutrinos on the turn-around radii is different from that of $f(R)$ gravity alone especially for the case that the $f(R)$ gravity 
is not strong enough to prostrate the effect of free streaming massive neutrinos. 

We make the same analysis but of the cluster halos identified at two different redshifts, $z=0.2$ and $0.4$ to see how $P(\alpha)$ changes with redshifts, 
the results of which are shown in Figures \ref{fig:cpro2} and \ref{fig:cpro3}, respectively. As can be seen, $P(\alpha)$, diminishes more rapidly with $\alpha$ 
at higher redshifts. While the statistical significance of the difference in $P(\alpha=4)$ between the $\frff$ and the other three models are quite 
robust against the redshift variation, that in $P(\alpha\ge 8)$ between the $\lcdm$ and $\frss$ models drops to $2.3\sigma_{\Delta P}$ at $z=0.2$ and 
to negligible level at $z=0.4$.  The low abundance of the cluster halos at higher redshifts contribute to the large uncertainties in $P(\alpha)$, rendering it 
inconclusive whether or not the strongest degeneracy between $\lcdm$ and $\frss$ can be broken by the turn-around radii of the cluster halos at $z\ge 0.2$. 

\section{Summary and Conclusion}\label{sec:con}

We have numerically demonstrated that the turn-around radii of cluster halos can in principle be useful to detect the effect of $f(R)$ gravity attenuated 
by the presence of massive neutrinos.  
The samples of the cluster halos with $M_{v}\ge 4.0\times 10^{14}\munit$ at $z=0$ were obtained from the {\small DUSTGRAIN}-{\it pathfinder} simulations \citep{gio-etal19} 
performed for four different cosmologies: the Planck $\lcdm$ and three $\nucdm$+$f(R)$ gravity models having different strength of fifth force and total neutrino mass: 
$\frs$, $\frss$, $\frff$, which were known to be degenerate with the $\lcdm$ model and with one another, yielding very similar conventional statistics \citep{bal-etal14,hag-etal19a}. 

\begin{figure}[ht]
\begin{center}
\plotone{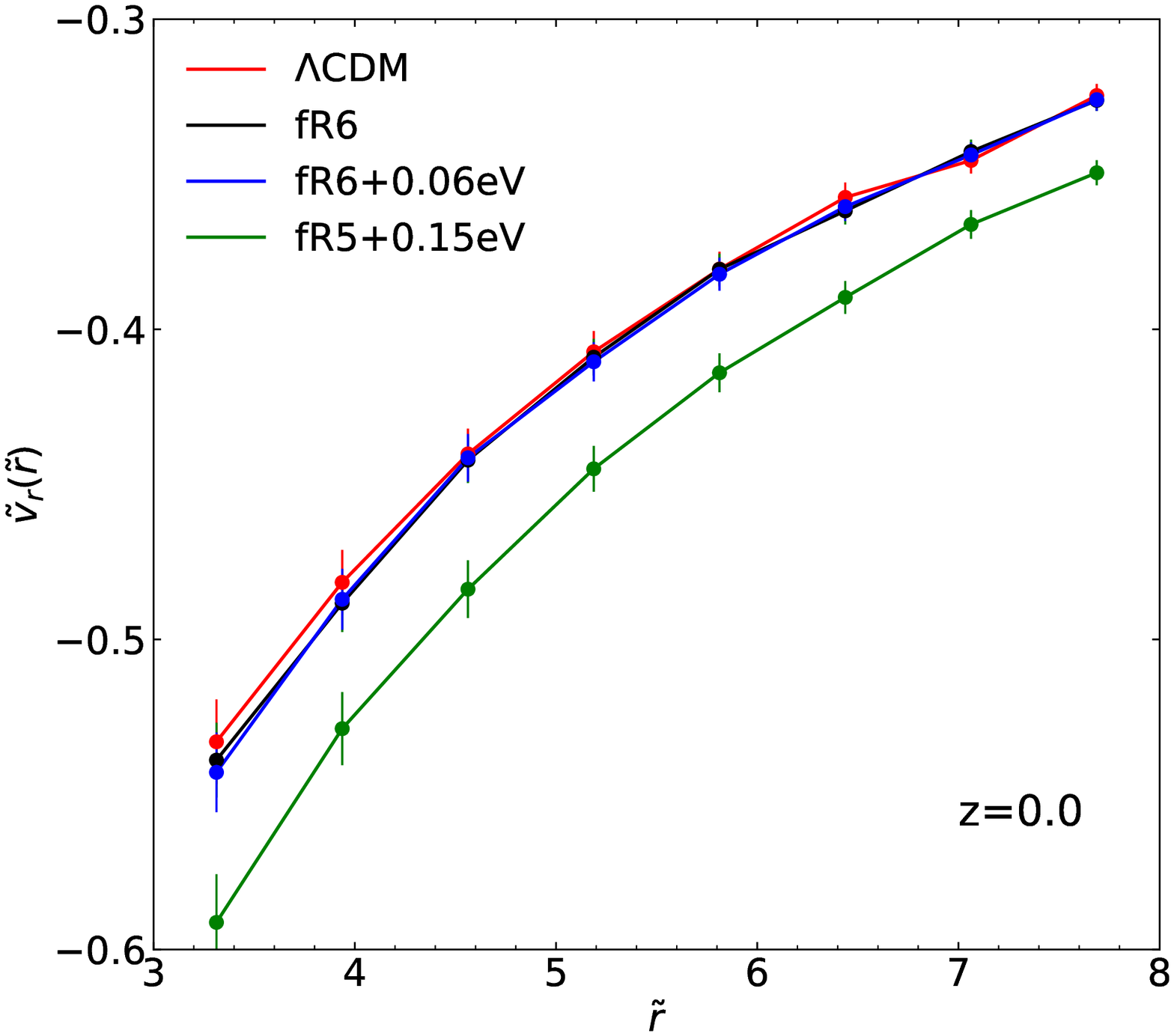}
\caption{Bound-zone velocity profiles around the cluster halos with $M_{v}\ge 4.05\times 10^{13}\,h^{-1}\,M_{\odot}$ 
at $z=0$ for four different cosmological models.}
\label{fig:vr}
\end{center}
\end{figure}

\begin{figure}[ht]
\begin{center}
\plotone{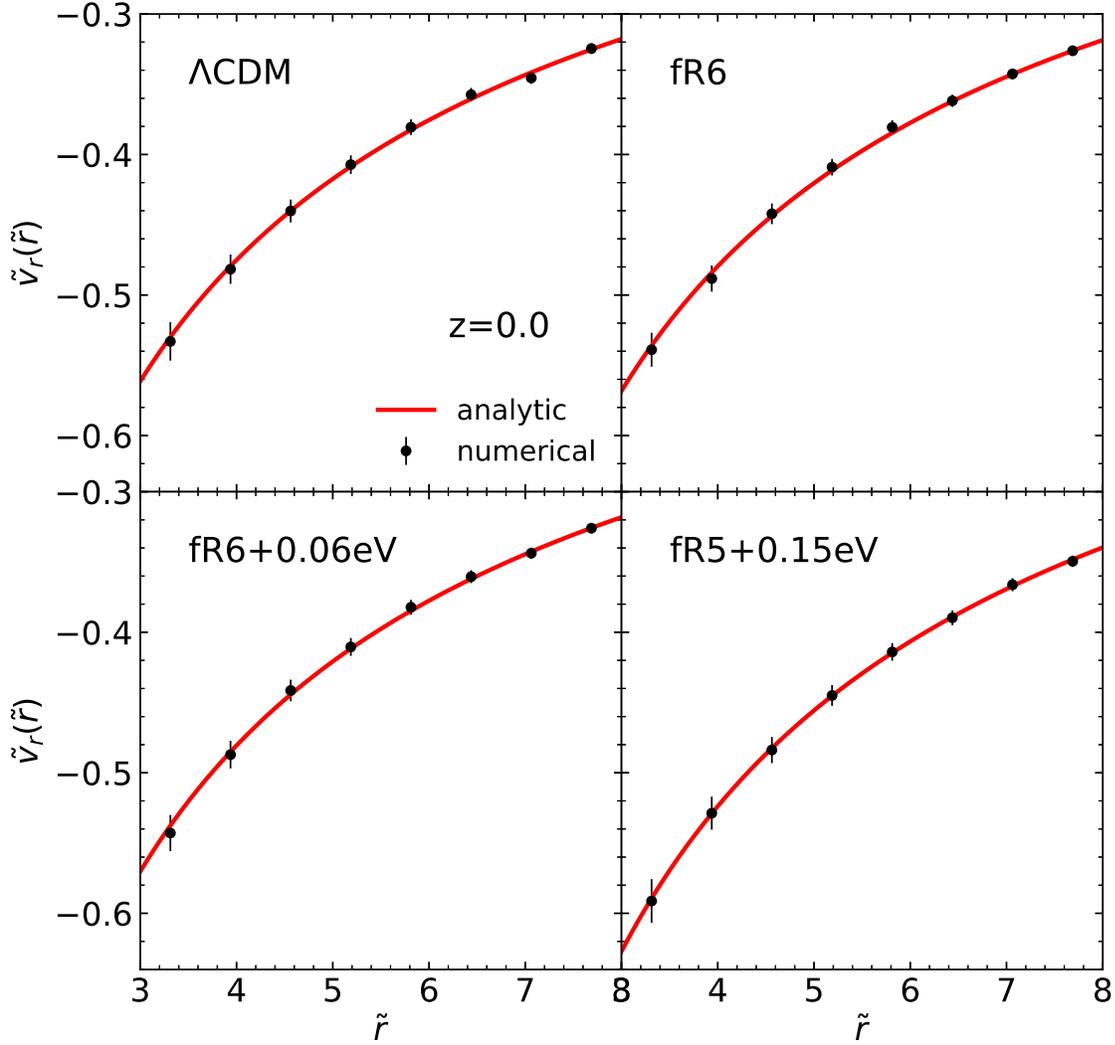}
\caption{Comparison of the numerically obtained bound-zone velocity profiles (black filled circles) with the best-fit 
analytic formula (red solid lines) at $z=0$.}
\label{fig:the}
\end{center}
\end{figure}

\begin{figure}[ht]
\begin{center}
\plotone{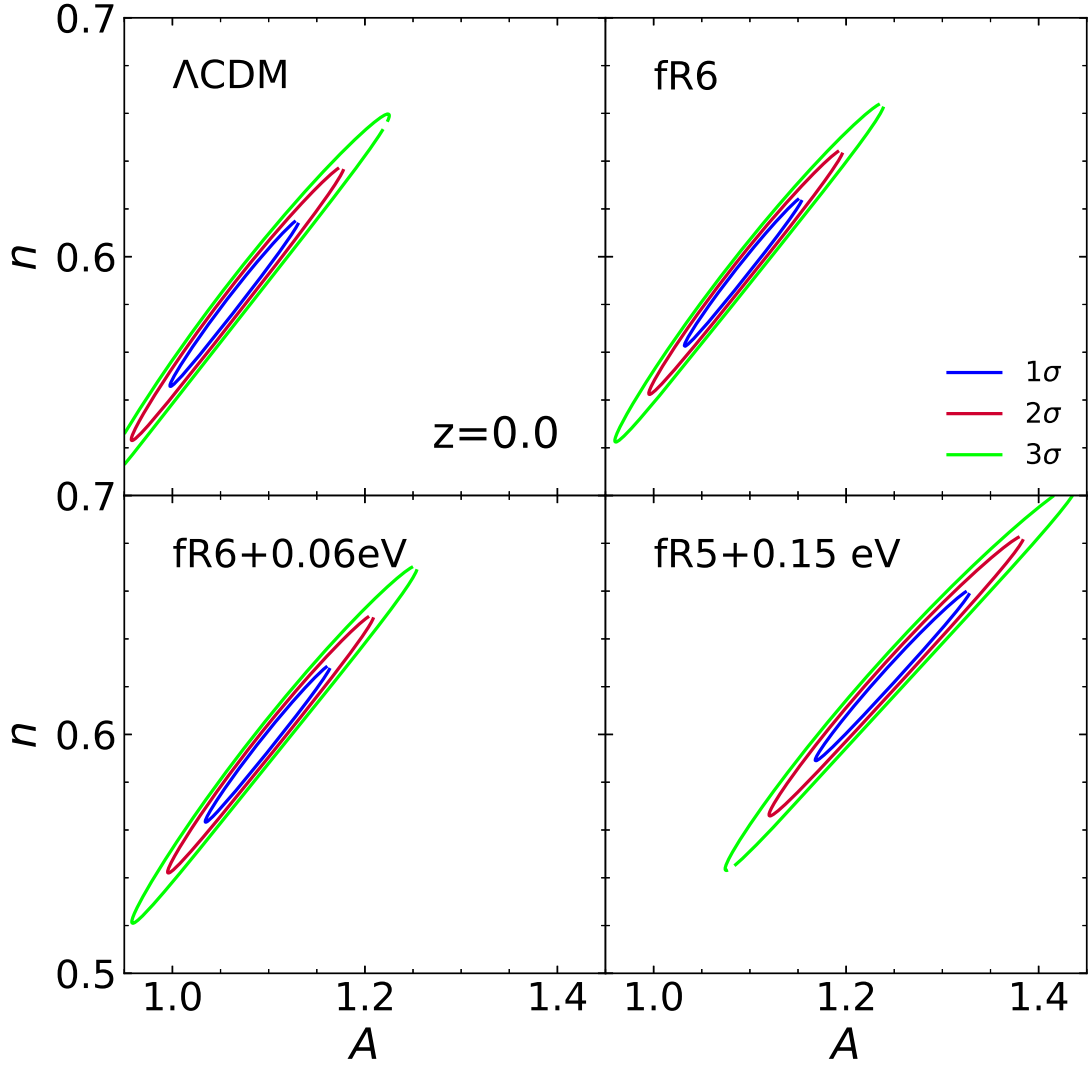}
\caption{$68\%$, $95\%$ and $99\%$ contours from the $\chi^{2}$-statistics for two parameters, $A$ and $n$, 
that characterize the analytic formula for the bound-zone velocity profiles, Equation (\ref{eqn:vr}), at $z=0$.}
\label{fig:chi2}
\end{center}
\end{figure}

For the determination of the turn-around radii, $r_{t}$, of the cluster halos at which the value of the peculiar velocity field becomes equal to the recession speed of the 
Hubble flow, we have employed the TRE (turn-around radius estimator) developed by \citet{lee-etal15}, which in turn utilizes the universal analytic formula for the peculiar 
velocity profile in the bound zone around the cluster halos, put forth by \citet{fal-etal14} for the $\lcdm$ case. 
Our comparison of the analytic formula with the average bound-zone velocity profile through the $\chi^{2}$ statistics has confirmed its validity  not only for the Planck 
$\lcdm$ but also for the three $\nucdm$+$f(R)$ gravity models. It has also revealed that the amplitudes and slopes of the bound-zone velocity profiles, quantified by its two adjustable 
parameters, significantly differ between the $\frff$ and the other three models (Figures \ref{fig:vr}-\ref{fig:chi2}). This result implies that the bound-zone velocities of the cluster halos must be 
much more susceptible to the presence of the strong fifth force than to that of massive neutrinos and thus that some statistics based on them may be useful to disentangle the former from 
the latter. 

\begin{figure}[ht]
\begin{center}
\plotone{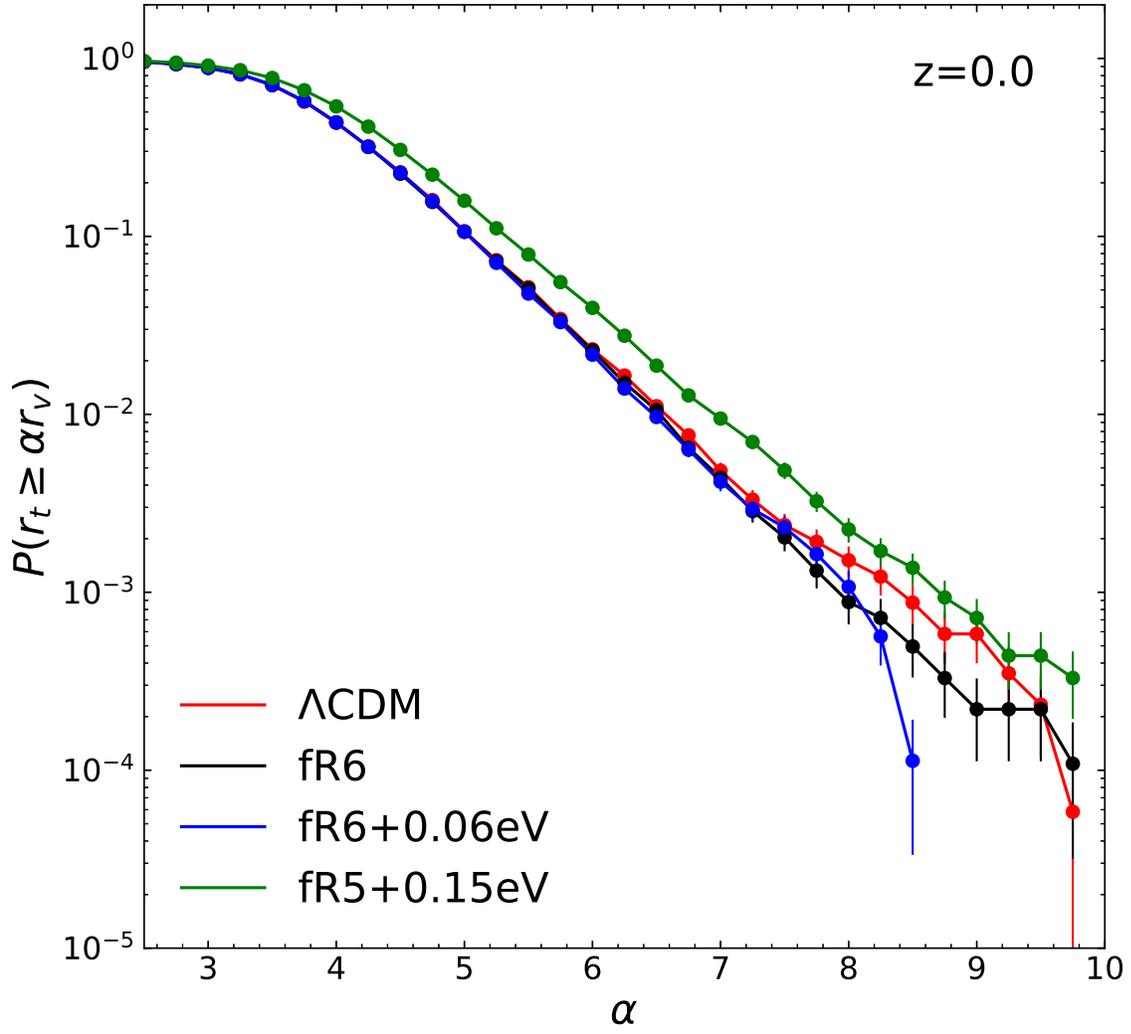}
\caption{Cumulative probability function of the ratios of the turn-around radii to the virial counterparts at $z=0$.}
\label{fig:cpro}
\end{center}
\end{figure}

\begin{figure}[ht]
\begin{center}
\plotone{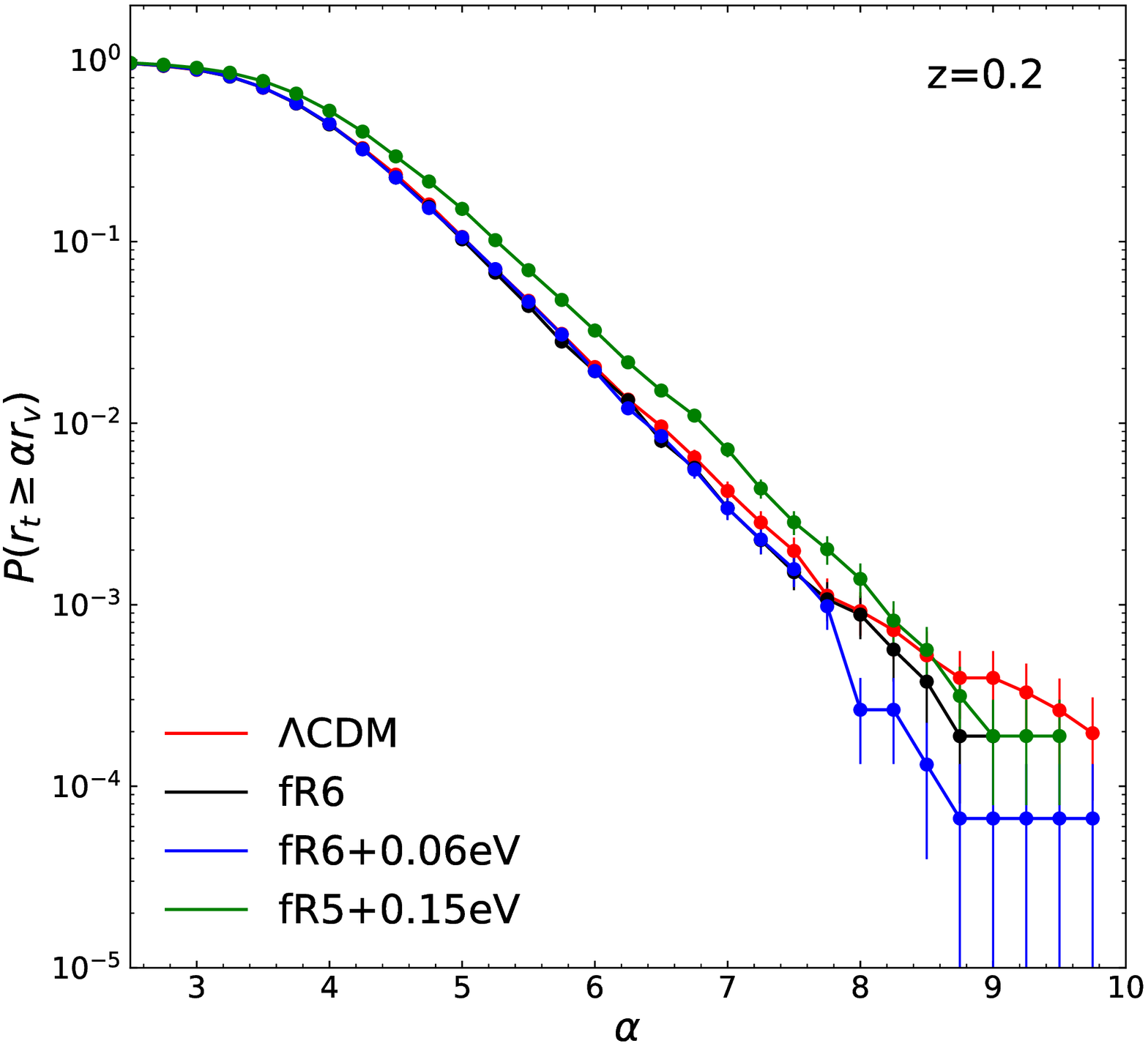}
\caption{Same as Figure \ref{fig:cpro} but at $z=0.2$.}
\label{fig:cpro2}
\end{center}
\end{figure}

\begin{figure}[ht]
\begin{center}
\plotone{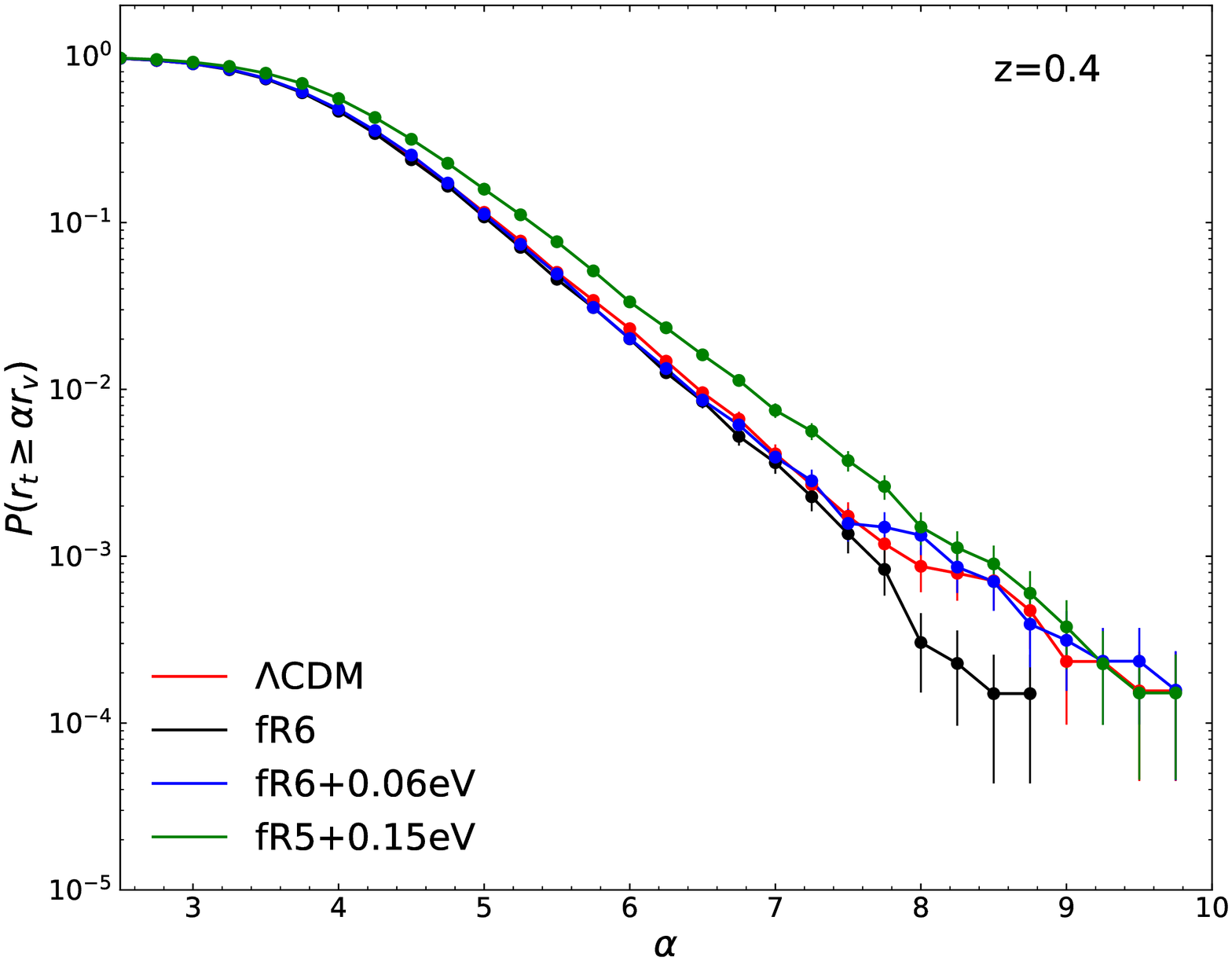}
\caption{Same as Figure \ref{fig:cpro} but at $z=0.4$.}
\label{fig:cpro3}
\end{center}
\end{figure}

With the turn-around radii of the cluster halos estimated by the application of the TRE to the bound-zone velocity profiles of individual cluster halos, we have determined 
the cumulative probability distributions of the turn-around to virial radius ratios,  $P(r_{t}\ge \alpha r_{v})$.  With the help of the bootstrap statistics, we have shown that the 
$\frff$ model can be plainly differentiated by $P(r_{t}/r_{v}\ge 4)$ from the other three cases, with statistical significance as high as $10\sigma_{\Delta P}$ (Figure \ref{fig:cpro}). 
We have also detected a $3.2\sigma_{\Delta P}$ difference in $P(r_{t}/r_{v}\ge 8.5)$ between the $\lcdm$ and $\frss$ models, in spite of the strongest degeneracy 
between the two cases. Yet, given the low value of $P(r_{t}/r_{v}\ge 8.5)\sim {\cal O}(10^{-4})$, it is not conclusive whether the $3.2\sigma_{\Delta P}$ difference in 
$P(r_{t}/r_{v}\ge 8.5)$ between the $\lcdm$ and $\frss$ models is a real signal or just a spurious one produced by the shot noise. A larger sample of the massive cluster 
halos with $r_{t}\ge 8 r_{v}$) will be required to confirm the statistical significance of the difference in $P(\alpha> 8)$ between the two models.
It has been also shown that the significance of the difference between the $\frff$ and the other three models is robust against the variation of the redshifts from $z=0.0$ to $0.4$ 
(Figures \ref{fig:cpro2}-\ref{fig:cpro3}). 

The advantage of using the turn-around radii estimated by the TRE to distinguish among the $\lcdm$ and $\nucdm$+$f(R)$ gravity models comes from the universality of 
the bound-zone velocity profile on which the TRE is based. As revealed by the previous works \citep{fal-etal14,lee16}, the shape of the bound-zone velocity profile 
is insensitive to the key cosmological parameters of the $\lcdm$ model. Thus, the variation of $\sigma_{8}$ and $\Omega_{m}$ in the $\lcdm$ model cannot 
produce the same effect on the turn-around radii as the $\nucdm$+$f(R)$ gravity. Furthermore, it does not require to track down the redshift evolution of $r_{t}$ 
unlike the previously suggested statistics as a possible discriminator of the $\nucdm$+$f(R)$ gravity model such as the size evolutions of galaxy voids,  
nonlinear growth rates and redshift distortions, evolution of the drifting average coefficient of the field cluster mass function, and 
high order weak lensing statistics \citep{pee-etal18,hag-etal19b,gio-etal19,wri-etal19,ryu-etal20,con-etal21}. 

It is, however, worth discussing the practical difficulties of our statistics. To detect a signal of the difference in $P(\alpha\ge 8)$ strong enough to distinguish among the degenerate 
models, what is required is to measure the turn-around radii of as many galaxy clusters as possible in the local universe. However, as shown in \citet{lee18} and \citet{han-etal20}, 
the TRE is applicable only to those isolated galaxy clusters whose bound zone neighbor galaxies exhibit very high degree of anisotropy in their spatial distributions. 
Due to this limitation of the TRE,  it would be difficult to obtain a large sample of the galaxy clusters with their turn-around radii measurable without information on the 
peculiar velocity field.  Notwithstanding, we expect that the large peculiar velocity dataset available from the future galaxy surveys like The Large Synoptic Survey Telescope (LSST) 
survey \citep{lsst} should allow us to directly measure the turn-around radii of almost all of the galaxy clusters at low-redshifts $z\le 0.2$, making our statistics based on the 
turn-around radii to be practically useful as a powerful discriminator of the $\nucdm$+$f(R)$ gravity models. 

\acknowledgments

JL acknowledges the support by Basic Science Research Program through the National Research Foundation (NRF) of Korea 
funded by the Ministry of Education (No.2019R1A2C1083855). JL thanks S.Ryu for having been helpful in obtaining the rockstar halo catalogs from 
the Dustgrain-Pathfinder simulations. MB acknowledges support by the project "Combining Cosmic Microwave Background and Large Scale 
Structure data: an Integrated Approach for Addressing Fundamental Questions in Cosmology", funded by the MIUR Progetti di Ricerca di Rilevante 
Interesse Nazionale (PRIN) Bando 2017 - grant 2017YJYZAH.  MB also acknowledges the use of computational resources from the parallel computing 
cluster of the Open Physics Hub at the Physics and Astronomy Department in Bologna (\url{https://site.unibo.it/openphysicshub/en}) .

\clearpage
\clearpage
\clearpage
\clearpage
\clearpage
\clearpage
\clearpage
\begin{deluxetable}{ccccccc}
\tablewidth{0pt}
\tablecaption{Best Parameters and Probabilities of $r_{t}\ge \alpha r_{v}$ at $z=0$}
\setlength{\tabcolsep}{3mm}
\tablehead{Cosmology & $N_{h}$ & $A$ & $n$ & $P(\alpha=4)$ & $P(\alpha=8.5)$ \\
& & & & ($10^{-2})$ & ($10^{-3}$)}
\startdata
$\lcdm$ & $18188$  & $1.061\pm 0.045$ & $0.580\pm 0.023$ & $43.50\pm 0.37$ & $0.87\pm 0.22$ \\
fR6	& $19317$  & $1.086\pm 0.041$ & $0.590\pm 0.021$ & $43.69\pm 0.36$ & $0.50\pm 0.16$\\
fR6+$0.06\ev$	&  $18811$ &  $1.096\pm 0.043$ & $0.595\pm 0.021$  & $43.61\pm 0.37$ & $0.11\pm 0.08$\\
fR5+$0.15\ev$ & $19291$  & $1.245\pm 0.053$ & $0.625\pm 0.035$ & $53.74\pm 0.36$ & $1.37\pm 0.28$\\
\enddata
\label{tab:data}
\end{deluxetable}

\end{document}